\documentclass[runningheads]{llncs}
\usepackage[T1]{fontenc}
\usepackage{comment}
\usepackage{graphicx}
\begin{document}
\title{Open-source tool for Airway Segmentation in Computed Tomography using 2.5D Modified EfficientDet: Contribution to the ATM22 Challenge}
%
%
\author{Diedre Carmo\inst{1}\orcidID{0000-0002-5922-9120} \and
Leticia Rittner\inst{1}\orcidID{0000-0001-8182-5554} \and
Roberto Lotufo\inst{1}\orcidID{0000-0002-5652-0852}}
\authorrunning{D. Carmo et al.}
%
\institute{Medical Imaging Computing Lab, University of Campinas, Campinas, SP, Brazil\\ 
\email{diedre@dca.fee.unicamp.br}\\
\url{https://miclab.fee.unicamp.br/}}
\maketitle              
\begin{abstract}
Airway segmentation in computed tomography images can be used to analyze pulmonary diseases, however manual segmentation is labor intensive and relies on expert knowledge. This manuscript details our contribution to MICCAI's 2022 Airway Tree Modelling challenge, a competition of fully automated methods for airway segmentation. We employed a  previously developed deep learning architecture based on a modified EfficientDet (MEDSeg), training from scratch for binary segmentation of the airway using the provided annotations. Our method achived 90.72 Dice in internal validation, 95.52 Dice on external validation and 93.49 Dice on the final test phase, while not being specifically designed or tuned for airway segmentation. Open source code and a pip package for predictions with our model and trained weights are in~\textbf{\url{https://github.com/MICLab-Unicamp/medseg}}.

\keywords{airway segmentation  \and deep learning \and airway tree modelling challenge \and tool \and open source}
\end{abstract}
%
%
%

\section{Introduction}

This short paper is a description of the data and methodology involved on our submission to Airway Tree Modelling 2022 (ATM22). ATM22 advances the availability of annotated airway segmentation data, with an order of magnitude more data than older challenges such as EXACT09~\cite{lo2012extraction}. Our contribution to the challenge consists of employing our Modified EfficientDet Segmentation (MEDSeg) model~\cite{carmo2021multitasking}, directly in a 2.5D axial training, only in data from the challenge. No pre-training is involved, and we did not tune the training parameters to this specific dataset with any ablation.

The rest of this paper is structured as follows: Section~\ref{sec:data} will go over a summary of the involved data, Section~\ref{sec:method} will explain the method and experiment parameters, Section~\ref{sec:results} presentes quantitative and qualitative results of our training and Section~\ref{sec:conclusion} discuss and conclude our contribution.

\section{Data}
\label{sec:data}

The challenge contains 500 computed tomography (CT) scans, 300 for training, 50 for external validation and 150 for testing, collected from multi-sites. The scans were collected from the public LIDC-IDRI dataset~\cite{armato2011lung} and the Shanghai Chest hospital. Each CT scan is semi-automatically annotated, with initial segmentations from a deep learning model carefully delineated and double-checked by three radiologists with more than five years of professional experience to acquire the final refined airway tree structure~\cite{qin2019airwaynet,zhang2021fda,zheng2021alleviating,yu2022break}.

One scan was removed due to incorrect labeling, as requested by the challenge organizers. The training set was split into 20\% (60) for internal validation and 80\% (239) for training. The 50 official validation and 150 testing scans are the external validation and test sets, with metrics calculated by the challenge organizers. Preprocessing of volumes include Hounsfield Intensity clipping to the $[-1024, 600]$ range and subsequent intensity normalization to the $[0, 1]$ range. For training, we only use 2.5D axial slices containing some kind of annotation, which resulted in 103623 slices. Note that 2.5D refers to including both neighbor slices to the central labeled slice, resulting in a 3 channel input for the network.

\section{Method}
\label{sec:method}
\begin{figure}[ht]
\centering
\includegraphics[width=\textwidth]{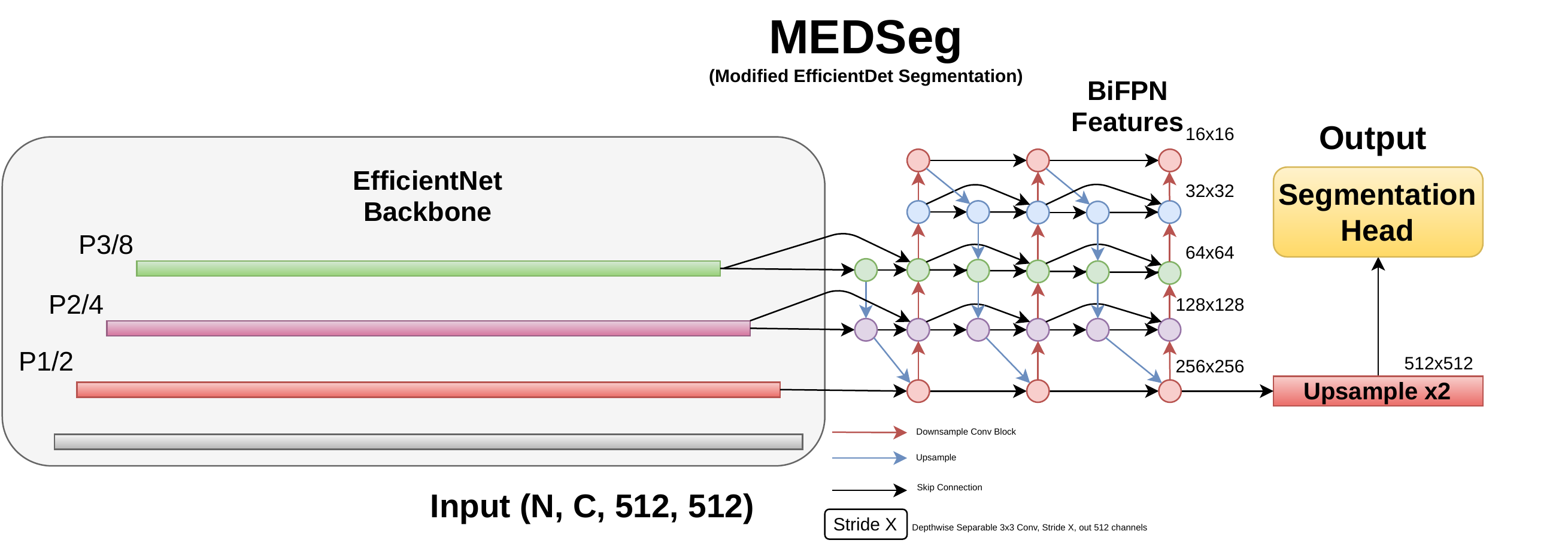}
\caption{Modified EfficientDet segmentation architecture (MEDSeg), where feature maps from EfficientNet are used as a BiFPN input, with padding of BiFPN operations for support of any input size, and the transposed upsample before the segmentation head. The segmentation head is composed of repeated blocks of separable convolutions, batch norm, and swish activations.}
\label{fig:edet}
\end{figure}

Our method is named Modified EfficientDet Segmentation (MEDSeg)~\cite{carmo2021multitasking}. MEDSeg is a novel take on EfficientDet~\cite{tan2020efficientdet}, a 2D natural image detection network. Starting from the original architecture, firstly we added padding for the spatial alignment of feature maps of the Bi-directional Feature Pyramid Network (BiFPN) downsampling and upsampling operations. This was necessary to be able to support any input spatial resolution, including odd shapes. The feature levels (P) that are used from EfficientNet were also changed, with the goal to access initial features half the size of the original image. Instead of using the P3, P4, P5, P6, and P7 features as in the original paper, we used P1, P2, and P3. This resulted in a final BiFPN output that is also half the size of the original image. To bring this feature representation to the size of the input image, we used simple bilinear upsampling. This representation is then fed to the segmentation head. Three blocks of depthwise convolutions~\cite{chollet2017xception} compose the segmentation head, batch normalization~\cite{ioffe2015batch}, and swish~\cite{ramachandran2017searching}, followed by a final convolution for channel reduction to the number of classes (Fig.~\ref{fig:edet}).

Axial 2.5D (3-channel) slices used as input in training are augmented through random 256x256 crop. Training of this method used the AdamW~\cite{loshchilov2017decoupled} optimizer, with initial learning rate of $1e-4$, exponential learning rate decay of $0.985$, weight decay of $1e-5$. The 2D output of the network is a sigmoid activation optimized through Dice Loss~\cite{sudre2017generalised}. Final volumetric segmentations are generated through stacking of 2D (1-channel) predictions, threshold of 0.5, and subsequent largest component extraction post-processing. For this application of MEDSeg, we did not change any parameters previously optimized for COVID-19 findings segmentation~\cite{carmo2021multitasking}, and just directly trained once in the provided data. The hidden evaluation of the challenge runners used metrics others than Dice, including Tree length detected rate (TD) and the branches detected rate (BD)~\cite{lo2012extraction}. We did not take into consideration these metrics while training MEDSeg.

\section{Results}
\label{sec:results}

\begin{figure}[ht]
\centering
\includegraphics[width=\textwidth]{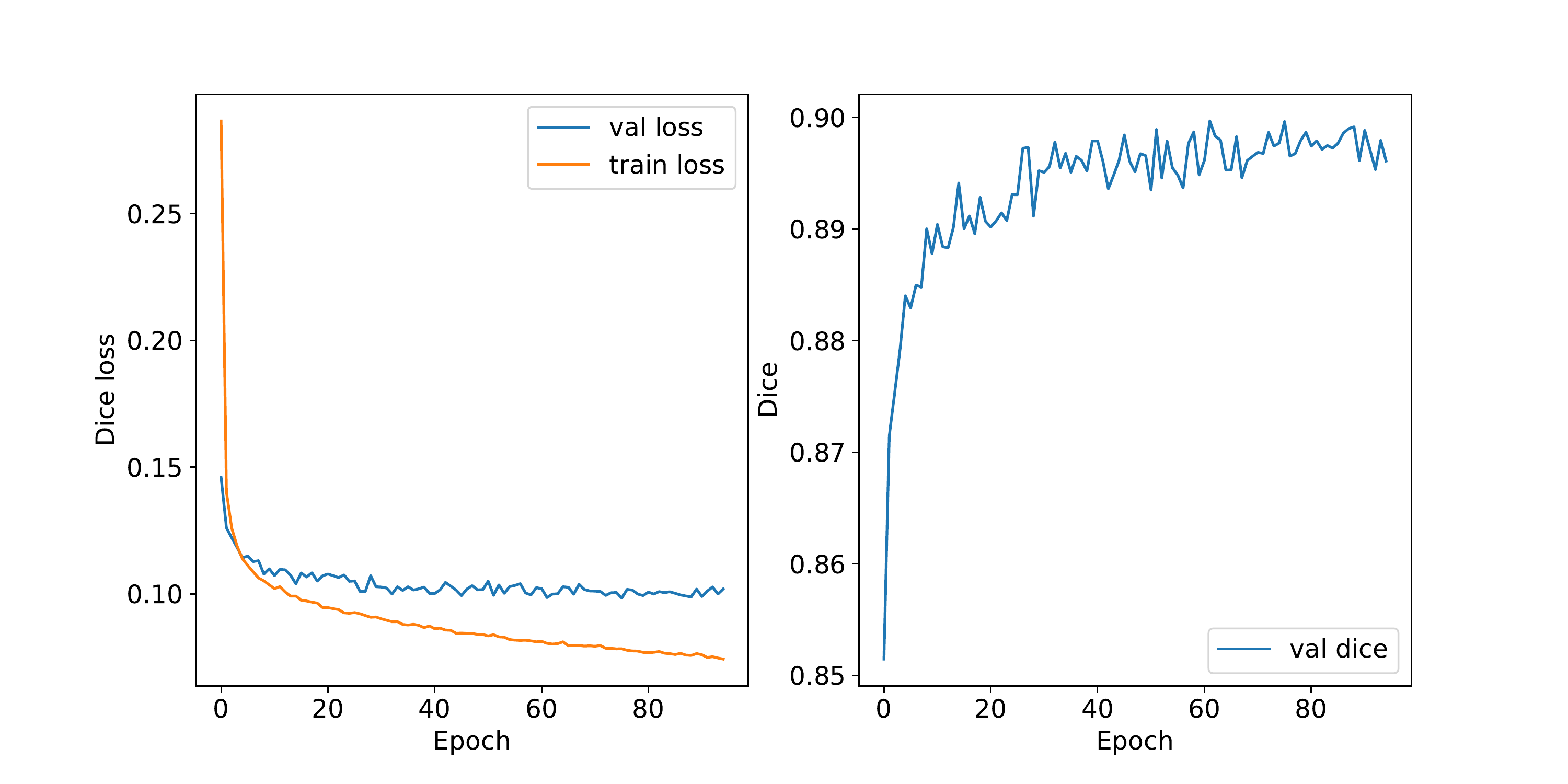}
\caption{Training and validation loss curves and validation Dice curve, per epoch.}
\label{fig:plot}
\end{figure}

Training using a 3080 Ti GPU took 84 hours, with a total of 95 epochs (an epoch corresponds to having seen a random crop from every axial slice in the training set). Progress of training (Dice) loss and internal validation loss and mean 2D Dice can be seen in Figure~\ref{fig:plot}, with signs of a beginning of overfitting. The lowest validation loss was reached in epoch 75, and corresponds to the model used for the following evaluations.

Table~\ref{tab:res} showcases 3D metrics for internal and external validation and the final test. We included False Negative Error (FNE) and False Positive Error (FPE) alongside our internal Dice evaluation. External validation and test were performed by the challenge runners.

\begin{table}[ht]
\centering
\caption{\label{tab:res}Results of our model in all three evaluation sets. External validation and test were performed by the challenge runners.}
\begin{tabular}{|l|l|l|l|l|l|}
\hline
\textbf{Evaluation Set} & \textbf{Dice (\%)}  & \textbf{FNE (\%)} & \textbf{FPE (\%)} & \textbf{TD (\%)} & \textbf{BD (\%)}\\ \hline
Internal Validation     & $90.72\pm4.13$      & $13.01\pm6.63$    & $4.92\pm2.49$     &      -           &      -          \\
External Validation     & $95.52\pm1.01$      & -                 &       -           & $82.85\pm5.93$   & $74.22\pm8.89$  \\
Test                    & $93.49$             & -                 & -                 &      $75.41$     &      $65.99$    \\\hline
\end{tabular}
\end{table}

\begin{figure}[ht]
\centering
\includegraphics[width=\textwidth]{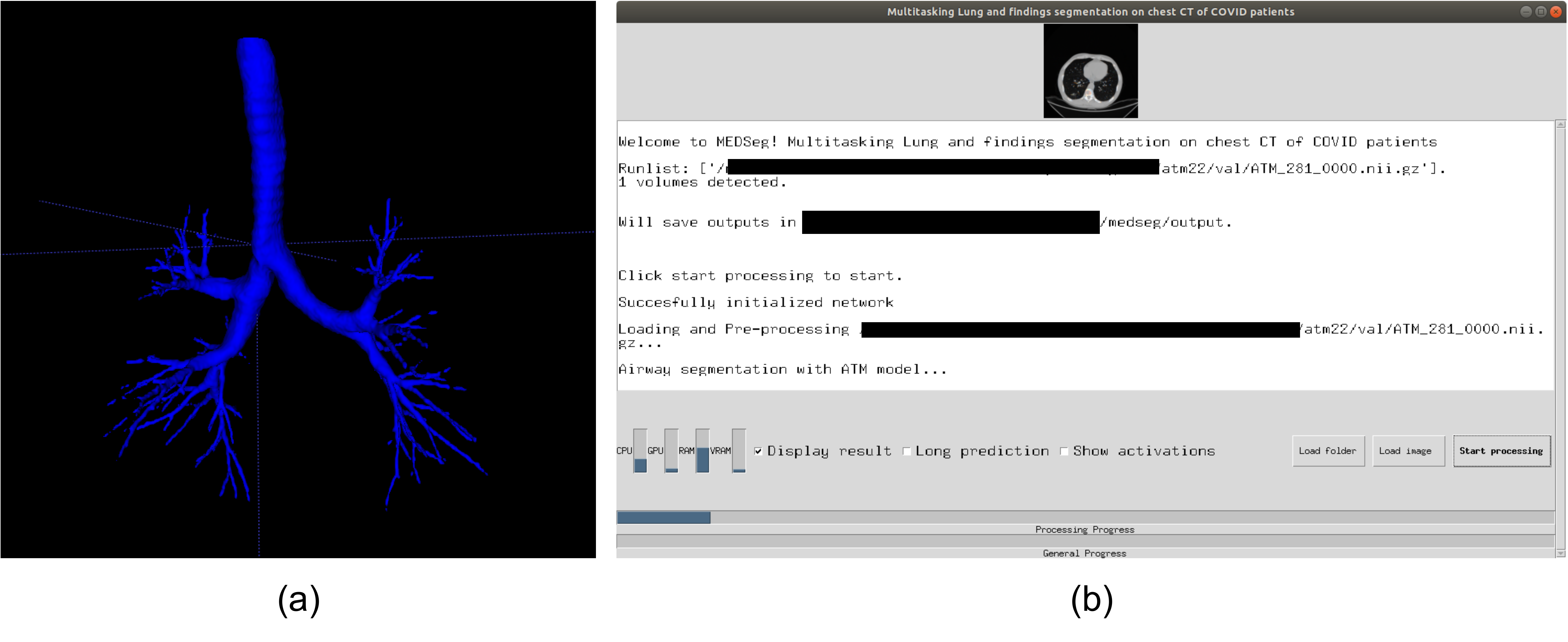}
\caption{(a) 3D rendering of the airway segmentation for a volume from a different dataset using ITK-Snap~\cite{py06nimg} and (b) screen capture of the GUI provided with the medseg pip package.}
\label{fig:see}
\end{figure}

In addition to quantitative metrics, we also deliver improvements to our graphical (Fig.~\ref{fig:see}) and command line user interface, providing a open source easy way to predict not only airway segmentations but lung and findings segmentation from our past work~\cite{carmo2021multitasking}, including output sheets with volumetric measurements and other statistics.

\section{Discussion and Conclusion}
\label{sec:conclusion}
Its interesting to note that a model that was not specifically designed for airway segmentation was able to achieve test Dice of 93.5 in airway segmentation. However, the challenge scoring took into consideration airway specific metrics that were not considered in our model. This shows how optimizing only for Dice can cause problems on fields where small portions of the segmented data, such as the correct prediction of airway branches, are more important than the bulk of the segmentation. Our contribution ended up being 6th place in the challenge validation leaderboard and 16th place in final testing for the challenge. We did not perform any hyperparameter tuning or ablation for this specific dataset, therefore future work might involve attempting multitasking for improved results and specific training and post-processing tuning to optimize metrics such as TD and BD for improved results. Easy to use command line interface, graphical user interface, pip package, Docker image, code and trained weights are available in~\textbf{\url{https://github.com/MICLab-Unicamp/medseg}}.

\section*{Acknowledgements} We thank grant \#2019/21964-4, São Paulo Research Foundation (FAPESP).

%
%
%

\bibliographystyle{splncs04}
\bibliography{bib}

\end{document}